\documentclass[prl,twocolumn,showpacs,superscriptaddress,aps]{revtex4}
\usepackage{amsfonts}
\usepackage{amsmath}
\usepackage{amssymb}
\usepackage{graphicx,color}
\usepackage{dcolumn}
\usepackage{times}

\DeclareMathOperator{\sgn}{sgn}

\begin{document}

\title{Light meets water in nonlocal media: Surface tension analogue in optics}

\author{Theodoros P. Horikis}
\affiliation{Department of Mathematics, University of Ioannina,
Ioannina 45110, Greece}

\author{Dimitrios J. Frantzeskakis}
\affiliation{Department of Physics, National and Kapodistrian University of Athens,
Panepistimiopolis, Zografos, Athens 15784, Greece}

\begin{abstract}

Shallow water wave phenomena find their analogue in optics through a nonlocal nonlinear
Schr\"{o}dinger (NLS) model in $(2+1)$-dimensions. We identify
an analogue of surface tension in optics, namely a single parameter depending
on the degree of nonlocality, %and the amplitude of the continuous-wave background,
which changes the sign of dispersion, much like surface tension does
in the shallow water wave problem. Using multiscale expansions, we reduce the NLS model
to a Kadomtsev-Petviashvilli (KP) equation, which is of the KPII (KPI)
type, for strong (weak) nonlocality.
%of either KPI or KPII type, depending on the magnitude of an effective negative
%surface tension. The latter,
%%Furthermore, this surface tension, the phenomenon that causes fluids
%%to minimize the area they occupy,
%is linked to the physical parameters of the original nonlocal system and it is thus shown that
%nonlocality is its direct analogue.
We demonstrate the emergence of robust optical antidark solitons
forming Y-, X- and H-shaped wave patterns, which are approximated by
colliding KPII line solitons, similar to those observed in shallow waters.
%
%We demonstrate
%how soliton solutions and their interaction patterns, as
%observed in shallow waters, may now be studied in nonlinear optics.

\end{abstract}


\maketitle

Many physically different
%subjects
contexts can be brought together through
%their
common modeling and
mathematical description.
%Perhaps the most
A common (and rather unlike)
example is water waves and nonlinear optics.
Two models
%systems
are inextricably linked with both subjects:
the universal Korteweg-de Vries (KdV) and nonlinear Schr\"odinger (NLS) equations \cite{MJA1}.
%
%Remarkable as these systems may be, for several physically relevant
%%contexts
%settings their standard form
%%turns out to be
%provide an oversimplified description as,
%% it cannot model,
%e.g., in the case of
%%for example,
%higher dimensionality; for instance, the Kadomtsev-Petviashvilli (KP) equation
%%is used as a generalization of the
%generalizes KdV to two spatial dimensions.
%
Furthermore, these
%systems
models can be reduced from one to the other \cite{zakharov},
thus suggesting that phenomena occurring in water waves
%will
may also exist in optics.
%In fact, in this work,
Here, using such reductions for a nonlocal NLS,
%model,
we find that surface tension --which causes fluids to minimize the area they occupy--
has a direct analogue in optics.
%equation.

Key to our findings are solitons, i.e.,
%solutions, the
robust localized waves that
%have always been a central element
play a key role in numerous studies in physics \cite{dp}, applied mathematics \cite{BlackBook}
and engineering \cite{hasko}. A unique property of solitons is that they feature
a particle-like character,
%which enables them to
i.e., they interact elastically, preserving
their shapes and velocities after colliding with each other. Such elastic collisions,
as well as pertinent emerging wave patterns,
can sometimes be observed even in everyday life. A predominant example is the one pertaining to
flat beaches: in such shallow water wave settings, two line solitons merging at proper
angles give rise to %two-dimensional wave
patterns of X-, H-, or Y-shaped waves, as well as other
more complicated nonlinear waveforms \cite{PRE}. All these
shallow water wave structures
%that are observed in shallow water
are actually exact analytical multi-dimensional line soliton solutions of the
Kadomtsev-Petviashvilli (KP) equation (which generalizes KdV to
%$(2+1)$-
two dimensions (2D) \cite{MJA1,BlackBook})
%-II (KPII) equation --
%which
%plays a key role
%is
of the KPII type -- a key model in the theory of shallow water
waves with weak surface tension \cite{MJA1}.
The relative equation with strong surface tension is referred to as KPI.

%It is the purpose of this work to
Here we show that such
%nonlinear
patterns can also be observed in a quite different physical setting,
%namely
i.e., the one related to optical beam propagation in media with a spatially nonlocal
defocusing nonlinearity. Such media include thermal nonlinear optical
media \cite{rot,krol1}, partially ionized plasmas \cite{litvak,plasma}, nematic liquid crystals \cite{ass0,ass1}, and
dipolar bosonic quantum gases \cite{dipolar}.
It is shown that approximate solutions of the nonlocal NLS model
satisfy, at proper scales, equations that appear in the context of water waves:
%an equation of the
a Boussinesq or Benney-Luke (BL) \cite{BL},
% type,
as well as a KP equation.
For a relatively strong (weak) nonlocality, or background amplitude, %or nonlocality,
the latter is found to be a KPII (KPI), similarly to the water wave problem,
where a KPII (KPI) results in the case of small (large)
surface tension \cite{MJA1}. Our results thus suggest an analogue of
surface tension in optics. Direct numerical simulations show that approximate antidark line
soliton solutions of the nonlocal NLS, constructed from the KPII line soliton solutions,
form patterns
%that are
observable in shallow water \cite{PRE}. Pertinent Y-, X-, or H-wave patterns may
%in principle
be realized in an experimental setup similar to the one
used for the observation of antidark solitons
%non-instantaneous optical media
\cite{segev2}.

The evolution of optical beams in nonlinear defocusing media is governed by the following
paraxial wave equation (cf. Ref.~\cite{kivshar_book} for
derivation and relevant adimensionalizations):
\begin{align}
i u_{t} + \frac{1}{2} \Delta u-nu =0,
\label{NLS1}
\end{align}
where subscripts denote partial derivatives, $u$ is the complex electric field envelope,
$\Delta\equiv\partial_x^2+\partial_y^2$ is the transverse Laplacian, and real function
$n$ denotes the nonlinear, generally nonlocal, medium response. For instance, in optics, $n$ is
the nonlinear change of the refractive index depending on the intensity $I=|u|^2$
\cite{rot,krol1}, in plasmas is the relative electron temperature perturbation \cite{litvak,plasma},
in liquid crystals is the optically induced angle perturbation \cite{ass0,ass1}, and so on.
Here, we consider that $n$ obeys the following diffusion-type equation:
\begin{align}
d^2\Delta n - n + |u|^2 = 0,
\label{NLS2}
\end{align}
where $d$ is a %characteristic
spatial scale (setting the diffusion length) that
measures the degree of nonlocality. Note that for $d = 0$, Eqs.~(\ref{NLS1})-(\ref{NLS2})
reduce to the %standard
defocusing 2D NLS equation
%in $(2+1)$-dimensions
\cite{kivshar_book}.
Importantly, Eqs.~(\ref{NLS1})-(\ref{NLS2}), the nonlocal NLS model of principal interest herein,
has been used satisfactorily to model experiments on liquid solutions
exhibiting thermal nonlinearities \cite{liq1,liq3}, while it has also been used in studies
of plasmas \cite{litvak,plasma} and nematic liquid crystals \cite{ass0,ass1}.

The steady-state
%homogeneous
solution of Eqs.~(\ref{NLS1})-(\ref{NLS2}) is composed by the continuous
wave (cw), $u=u_0\exp(-i|u_0|^2 t)$ %(where $u_0$ is an arbitrary
($u_0$ being an arbitrary complex constant),
and the constant function $n=|u_0|^2$. Considering small perturbations of this solution
behaving like $\exp[i(\boldsymbol{k}
\cdot \boldsymbol{r} -\omega t)]$, with $\boldsymbol{r}=(x,~y)$, we find that the
perturbations' wavevector $\boldsymbol{k}=(k_x,~k_y)$ and frequency $\omega$ obey the dispersion
relation:
\begin{equation}
\omega^2=|\boldsymbol{k}|^2 C^2 \left(1+d^2|\boldsymbol{k}|^2\right)^{-1}
+(1/4)|\boldsymbol{k}|^4,
\label{dr}
\end{equation}
where $C^2=|u_0|^2$ is the wave velocity. Here it is important to observe the following. First,
%Evidently,
since $\omega \in \mathbb{R}$~$\forall ~\boldsymbol{k}$,
%which indicates that
the steady-state
%homogeneous
solution is modulationally stable.
Second, in the long-wavelength limit ($|\boldsymbol{k}|^2 \ll1$),
Eq.~(\ref{dr}) becomes $\omega^2 \approx |\boldsymbol{k}|^2 C^2+(1/4) \alpha |\boldsymbol{k}|^4$,
where $\alpha = 1-4d^2|u_0|^2$.
%
%\begin{equation}
%\alpha = 1-4d^2|u_0|^2.
%\label{alpha}
%\end{equation}
%%
%
This approximate dispersion relation features a striking similarity to the
corresponding (approximate) one for shallow water waves, namely \cite{MJA1}:
$\omega^2 \approx |\boldsymbol{k}|^2 c_0^2+(1/3) (3\hat{T}-1) c_0^2 h^2 |\boldsymbol{k}|^4$,
where $c_0^2 = gh$ is the velocity, $g$ is the acceleration of gravity,
$h$ the depth of water at rest, and $\hat{T}=T/(\rho g h^2)$, with $\rho$ being the density
and $T$ the surface tension. Comparing these dispersion relations, the
following correspondence is identified: $3\hat{T} \rightarrow 4d^2|u_0|^2$,
implying that there exists a surface tension analogue in our problem,
$\propto d^2|u_0|^2$.
%by the nonlocality parameter $d$ (and the cw intensity $|u_0|^2$).
This effective surface tension is negative, as is also implied by the fact that the term
$\propto d$ in the Hamiltonian
$\mathcal{H}=(1/2) \int_{\mathbb{R}^2}
\left(|\boldsymbol{\nabla} u|^2-d^2(\boldsymbol{\nabla} n)^2-n^2+2n|u|^2\right)
{\rm d}\boldsymbol{r}$ of Eqs.~(\ref{NLS1})-(\ref{NLS2}) decreases the
potential energy of the system, oppositely to the water wave case where %the
surface tension increases the respective potential energy \cite{zakh}.

These arguments can further be solidified by analyzing the fully nonlinear problem:
similarly to water waves \cite{MJA1}, we will derive KPI and KPII equations,
depending on the strength of the effective surface tension, i.e., the parameter
$\alpha$, which sets the dispersion coefficient in KP. This
can already be identified from the linear theory as follows.
Using $|\boldsymbol{k}|^2 =k_x^2+k_y^2$, the long-wavelength limit of Eq.~(\ref{dr})
%of the dispersion relation of Eqs.~(\ref{NLS1})-(\ref{NLS2})
reads: $\omega= \pm C k_x\left[1+(k_y/k_x)^2\right]^{1/2} \left[1+(\alpha/4C^2) k_x^2+
\mathcal{O}(k_y^2)\right]$, %where
with $\pm$ corresponding to right- and left-going waves.
%Then,
Assuming
$|k_y/k_x|\ll 1$ and $k_y^2 \sim \mathcal{O}(k_x^4)$, we find:
$(1/C)\omega k_x = \pm \left[ k_x^2 + (\alpha/4C^2) k_x^4 +(1/2)k_y^2\right]$.
%To this end,
Then, using
$\omega \rightarrow i\partial_t$, $k_{x,y} \rightarrow -i\partial_{x,y}$,
%it is found that yields
the linear PDE associated to this dispersion relation is:
$\partial_x [\pm q_t+ C q_x - (\alpha/8C)q_{xxx}] +(C/2)q_{yy}=0$.
This is a linear KP equation, with a dispersion coefficient depending
on the effective surface tension through $\alpha$, similarly to shallow
water waves, where the respective dispersion coefficient depends
on $\hat{T}$ \cite{MJA1}. %, as noted above.

To derive the full nonlinear version of the KP equation, we resort to multiple scales.
%To do so, we
%We can, thus,
We thus consider small-amplitude slowly-varying modulations of the steady state, and seek %for
solutions of Eqs.~(\ref{NLS1})-(\ref{NLS2}) in the form of the
%following
asymptotic expansions:
%
%\begin{eqnarray}
%u&=&u_0\sqrt{\rho}\exp\left[-i|u_0|^2 t + i \epsilon^{1/2}\Phi(X,Y,T)\right],
%\label{u} \\
%\rho&=& 1+\sum_{j=1}^{\infty}\epsilon^j \rho_j(X,Y,T),
%\label{rho} \\
%n&=&|u_0|^2 + \sum_{j=1}^{\infty}\epsilon^j n_j(X,Y,T),
%\label{n}
%\end{eqnarray}
%
\begin{eqnarray}
u&=&u_0\sqrt{\rho}\exp(-i|u_0|^2 t + i \epsilon^{1/2}\Phi),
\label{u} \\
\rho&=& 1+\sum_{j=1}^{\infty}\epsilon^j \rho_j, \quad
n=|u_0|^2 + \sum_{j=1}^{\infty}\epsilon^j n_j,
\label{n}
\end{eqnarray}
where $0<\epsilon \ll1$ is a formal small parameter, while %the
phase $\Phi$ and amplitudes $\rho_j$ and $n_j$ are unknown real
functions of the slow variables $X=\epsilon^{1/2}x$, $Y=\epsilon^{1/2}y$
and $T=\epsilon^{1/2}t$. Substituting the expansions~(\ref{u})-(\ref{n}) into
Eqs.~(\ref{NLS1})-(\ref{NLS2}),
%and equating terms of the same order in $\epsilon$,
we obtain the
following results. First, the real part of Eq.~(\ref{NLS1}) and Eq.~(\ref{NLS2}) yield
%, respectively,
the leading-order equations, at $\mathcal{O}(\epsilon^{3/2})$ and
$\mathcal{O}(\epsilon)$:
\begin{eqnarray}
\rho_{1T} +\tilde{\Delta}\Phi=0, \quad n_1=|u_0|^2\rho_1,
\label{lead}
\end{eqnarray}
and the first-order equations, at $\mathcal{O}(\epsilon^{5/2})$ and $\mathcal{O}(\epsilon^2)$:
\begin{eqnarray}
\rho_{2T}+\tilde{\boldsymbol{\nabla}} \cdot (\rho_1 \tilde{\boldsymbol{\nabla}}\Phi)=0,
\quad
d^2 \tilde{\Delta}n_1-n_2 +|u_0|^2\rho_2=0,
\label{first}
\end{eqnarray}
connecting the amplitudes $\rho_{1,2}$ and $n_{1,2}$ with the phase $\Phi$; here, $\tilde{\Delta}
\equiv \partial_X^2 + \partial_Y^2$ and
$\tilde{\boldsymbol{\nabla}}\equiv (\partial_X,~\partial_Y)$.
Second, the imaginary part of Eq.~(\ref{NLS1}), combined with Eqs.~(\ref{lead})-(\ref{first}), yields:
%the following wave equation for $\Phi$:
%
\begin{eqnarray}
&&\Phi_{TT}-C^2\tilde{\Delta}\Phi + \epsilon
\left[ \frac{1}{4}\alpha \tilde{\Delta}^2\Phi
+ \frac{1}{2}\partial_T(\tilde{\boldsymbol{\nabla}}\Phi)^2
\right. \nonumber \\
&&\left.
+ \tilde{\boldsymbol{\nabla}} \cdot (\Phi_T \tilde{\boldsymbol{\nabla}}\Phi)
\right]=\mathcal{O}(\epsilon^2),
\label{B}
\end{eqnarray}
%
%where $C^2=|u_0|^2$ sets the wave velocity [which is identical to the one
%suggested by Eq.~(\ref{dr})], and
%%
%\begin{equation}
%\alpha = 1-4d^2|u_0|^2.
%\label{alpha}
%\end{equation}
%
%At leading-order, Eq.~(\ref{B}) is the standard wave equation.
%while
%at order $\mathcal{O}(\epsilon)$, the linear part of Eq.~(\ref{B}) corresponds to the dispersion
%relation~(\ref{dr}) for the small-amplitude linear waves of Eqs.~(\ref{NLS1})-(\ref{NLS2})
%propagating on top of the steady state with $|u|=|u_0|$ and $n=|u_0|^2$.
%
%On the other hand,
%The full
Equation~(\ref{B}) incorporates
%fourth
4th-order dispersion and quadratic
nonlinear terms, resembling the Boussinesq and Benney-Luke \cite{BL} equations,
%. These models, are used to
which describe bidirectional shallow water waves
%, in the framework of small-amplitude and long wave approximations
\cite{MJA1}. Similarly to
%Using a multiscale expansion method, similar to the one employed in
the water wave problem,
% \cite{MJA1},
we now use a multiscale expansion method
to derive the KP equation,
%which is obtained
under the additional assumptions of quasi-two-dimensionality and unidirectional propagation. In
particular, we introduce the asymptotic expansion $\Phi=\Phi_0+\epsilon \Phi_1 +\cdots$, where
%unknown
functions $\Phi_{\ell}$ ($\ell=0,1,\ldots$) depend on the variables $\xi=X-CT$, $\eta=X+CT$,
$\mathcal{Y}=\epsilon^{1/2}Y$, and $\mathcal{T}=\epsilon T$. Substituting this expansion into
Eq.~(\ref{B}), at the leading-order in $\epsilon$, we obtain the wave equation $\Phi_{0\xi
\eta}=0$, implying that $\Phi_0$ can be expressed as a superposition of a right-going wave,
$\Phi_0^{(R)}$, depending on $\xi$, and a left-going one, $\Phi_0^{(L)}$, depending on $\eta$,
namely:
\begin{equation}
\Phi_{0}=\Phi_0^{(R)}(\xi,\mathcal{Y},\mathcal{T})+\Phi_0^{(L)}(\eta,\mathcal{Y},\mathcal{T}).
\label{rl}
\end{equation}
In addition, at order $\mathcal{O}(\epsilon)$, we obtain the equation:
\begin{eqnarray}
&&4C^2\Phi_{1\xi\eta} = -C\left(\Phi_{0\xi\xi}^{(R)}\Phi_{0\eta}^{(L)}
-\Phi_{0\xi}^{(R)}\Phi_{0\eta\eta}^{(L)} \right) \nonumber \\
&&+\left[\partial_{\xi}
\left(-2C\Phi_{0\mathcal{T}}^{(R)} +\frac{\alpha}{4}\Phi_{0\xi\xi\xi}^{(R)}
-\frac{3C}{2}\Phi_{0\xi}^{(R)2}\right)
-C^2\Phi_{0\mathcal{Y}\mathcal{Y}}^{(R)}\right]
\nonumber \\
&&+ \left[\partial_{\eta}\left(2C\Phi_{0\mathcal{T}}^{(L)}
+\frac{\alpha}{4}\Phi_{0\tilde{\eta}\tilde{\eta}\tilde{\eta}}^{(L)}
+\frac{3C}{2}\Phi_{0\tilde{\eta}}^{(L)2}
\right)-C^2\Phi_{0\mathcal{Y}\mathcal{Y}}^{(L)}
\right].
\nonumber %\\
%\label{phicarsp}
\end{eqnarray}
When integrating this equation, secular terms arise from the square brackets,
which are functions of $\xi$ or $\eta$ alone, not both.
%Integration of Eq.~(\ref{phicarsp}) in $\xi$ or $\eta$, reveals that the terms in square
%brackets in the right-hand side are secular, as functions of $\xi$ or $\eta$ alone.
Removal of these secular terms leads to two uncoupled nonlinear evolution equations for
$\Phi_0^{(R)}$ and $\Phi_0^{(L)}$. %Furthermore, using the equation
Then, using $\Phi_T=-n_1$, obtained from the
leading-order part of Eq.~(\ref{B}) together with Eq.~(\ref{first}), it is found that the
amplitude $\rho_1$ can also be decomposed to a left- and a right-going wave, i.e., $\rho_1 =
\rho_1^{(R)}+\rho_1^{(L)}$, which satisfy the following KP equations:
%%
%\begin{eqnarray}
%\partial_{\mathcal{X}}\left(\pm \rho_{1\mathcal{T}}^{(R,L)}
%-\frac{\alpha}{8C}\rho_{1\mathcal{X}\mathcal{X}\mathcal{X}}^{(R,L)}
%+\frac{3C}{2}\rho_{1}^{(R,L)} \rho_{1\mathcal{X}}^{(R,L)}\right)
%\nonumber \\
%+\frac{C}{2}\rho_{1\mathcal{Y}\mathcal{Y}}^{(R,L)}=0,
%%\nonumber \\
%\label{KPa}
%\end{eqnarray}
%%
%
\begin{eqnarray}
\partial_{\mathcal{X}}\left(\pm \rho_{1\mathcal{T}}^{(R,L)}
-\frac{\alpha}{8C}\rho_{1\mathcal{X}\mathcal{X}\mathcal{X}}^{(R,L)}
+\frac{3C}{4}\rho_{1}^{(R,L)2} \right)
%\nonumber \\
+\frac{C}{2}\rho_{1\mathcal{Y}\mathcal{Y}}^{(R,L)}=0,
\nonumber
%\\
%\label{KPa}
\end{eqnarray}
where $\mathcal{X}=\xi$ ($\mathcal{X}=\eta$) for the right- (left-) going wave. Next,
%focusing on the case of the
for right-going waves, we use the
%introduce the
transformations
$ \mathcal{T} \rightarrow -(\alpha/8C)\mathcal{T}$,
$\mathcal{Y} \rightarrow \sqrt{3|\alpha|/4C^2}\mathcal{Y}$,
and
$\rho_1^{(R)}=-(\alpha/2C^2) U$,
and express KP
%Eq.~(\ref{KPa})
in its standard dimensionless form \cite{BlackBook,MJA1}:
\begin{eqnarray}
\partial_{\mathcal{X}}\left(U_{\mathcal{T}}+6UU_{\mathcal{X}}
+U_{\mathcal{X}\mathcal{X}\mathcal{X}}\right)
+3\sigma^2 U_{\mathcal{Y}\mathcal{Y}}=0,
\label{usKP}
\end{eqnarray}
where $\sigma^2=-\sgn{\alpha}=\sgn(4d^2|u_0|^2-1)$.
%
%parameter $\sigma^2$ is given by
%%
%\begin{equation}
%\sigma^2=-\sgn{\alpha}=\sgn(4d^2|u_0|^2-1).
%\label{sigma}
%\end{equation}
%
Importantly, Eq.~(\ref{usKP}) includes both versions of the KP equation, KPI and KPII
\cite{BlackBook}. Indeed, for $\sigma^2=1\Rightarrow\alpha<0$, i.e., for
%$|u_0|^2>1/(4d^2)$ or $d^2>1/(2|u_0|)^2$,
$4d^2 |u_0|^2>1$,
Eq.~(\ref{usKP}) is a KPII equation; on the other hand, for
$\sigma^2=-1\Rightarrow\alpha>0$, i.e.,
%$|u_0|^2<1/(4d^2)$ or $d^2<1/(2|u_0|)^2$,
$4d^2 |u_0|^2<1$, Eq.~(\ref{usKP}) is a KPI equation. Thus,
%
%for a fixed degree of nonlocality $d^2$,
%a larger (smaller) background amplitude
%$|u_0|$, as defined by the sign of $\alpha$, corresponds to KPII (KPI); similarly,
for a fixed cw intensity $|u_0|^2$, a strong (weak) nonlocality $d^2$,
as defined by the above inequalities,
%regimes of $d^2$,
corresponds to KPII (KPI); the same holds for a fixed degree of nonlocality $d^2$,
and a larger (smaller) cw intensity $|u_0|^2$.
Thus, both our linear and nonlinear analysis establishes a ``homeomorphism''
between optics and shallow water waves: in this latter context, weak surface tension (typical for
water waves) corresponds to $\sigma^2=1$ in Eq.~(\ref{usKP}) (i.e., to KPII), while strong surface tension is
pertinent to $\sigma^2=-1$ (i.e., to KPI) \cite{MJA1,BlackBook}.
%
%, where KPII (KPI) is derived for weak (strong) surface
%tension \cite{MJA1,BlackBook}.

%Notice that in the local limit of $d=0$, the asymptotic analysis leads
%only to the KPI model, in which line solitons are unstable; this fact was used to investigate
%self-focusing and transverse instability
%of plane dark solitons of the defocusing NLS equation \cite{peli1,pelirev}. Notably, the same
%parameter $\alpha$ has also been shown to distinguish solutions in 1D \cite{tph} and radially
%symmetric \cite{djfth} systems, where ring solitons were found.

%Here, it should be highlighted that the existence of the above regimes resembles the situation
%occurring in shallow water. In this context, weak surface tension
%corresponds to $\sigma^2=1$ in Eq.~(\ref{usKP}) (i.e., KPII), while strong surface tension is
%pertinent to $\sigma^2=-1$ (i.e., KPI). Thus, there exists an immediate connection between the
%original problem with the one of shallow water waves: relatively large (small) background amplitude
%or nonlocality corresponds to weak (strong) surface tension, leading to KPII (KPI).
%
Based on the above analysis, we now
%it is now straightforward to
utilize the exact solutions of the KP, Eq.~(\ref{usKP}), and
construct approximate solutions of the
original system of Eqs.~(\ref{NLS1})-(\ref{NLS2}); such solutions read:
%of the form:
%
\begin{align}
&u\approx u_0 \left(1-\epsilon \frac{\alpha}{2|u_0|^2}U \right)^{1/2} %\nonumber \\
%&\times
\exp\left(-i|u_0|^2 t\right)
%\right.
\nonumber \\
%&\left.
&\times\exp\left(\frac{i}{2} \alpha \epsilon^{-1/2} \int_0^\mathcal{T} U {\rm
d}\mathcal{T'}\right), \quad
%\label{apsol1} \\
n \approx |u_0|^2 - \frac{1}{2}\alpha U \label{apsol2}.
\end{align}
%
%Note that
Clearly, for $\alpha<0$ ($\alpha>0$), i.e., for solutions satisfying KPII (KPI),
%the approximate soliton solution for
$u$ in Eq.~(\ref{apsol2}) has the form of a hump (dip)
on top (off) of the cw background and is, thus, a antidark (dark) soliton.
%
%Here, we focus on the stable soliton solutions of the KP equations, namely the (antidark) line
%solitons of the KPII equation and the (dark) lump of KPI.
%

Notice that in the local limit of $d=0$ (i.e., $\alpha=1$), we
solely obtain the KPI model, in which line solitons are unstable: as was shown
in plasma physics and hydrodynamics \cite{zakhpr}, line solitons develop undulations and
eventually decay into lumps \cite{infeld}. In the same venue, but now in optics, the
asymptotic reduction of the defocusing 2D NLS to KPI \cite{kuz1,peli1},
and the instability of the line solitons of the latter, was used
to better understand the transverse instability of rectilinear
dark solitons: indeed, these structures also develop undulations and
eventually decay into vortex pairs \cite{peli1,pelirev}.

Here, we focus on the stable soliton solutions of the KP equations, namely the (antidark) line
solitons of the KPII equation and the (dark) lump of KPI. The one-line soliton solution,
travelling at an angle to the $\mathcal{Y}$-axis, is:
%
%\begin{subequations}
%\begin{gather}
%U(\mathcal{X},\mathcal{Y},\mathcal{T})= 2\kappa^2 {\rm sech}^2(\mathcal{Z}),
%\\
%\mathcal{Z} \equiv \kappa \left[ \mathcal{X}+ \lambda \mathcal{Y} -
%\left(4\kappa^2+3\lambda^2 \right) \mathcal{T} +\delta \right],
%\end{gather}
%\end{subequations}
%
\begin{equation}
U(\mathcal{X},\mathcal{Y},\mathcal{T})= 2\kappa^2 {\rm sech}^2(\mathcal{Z}),
\end{equation}
where $\mathcal{Z} \equiv \kappa \left[ \mathcal{X}+ \lambda \mathcal{Y} -
\left(4\kappa^2+3\lambda^2 \right) \mathcal{T} +\delta \right]$, with
$\kappa$, $\lambda$ and $\delta$ being free parameters. On the other hand,
the two-line soliton can be expressed in the following form:
%
%\begin{subequations}
%\begin{gather}
%U(\mathcal{X},\mathcal{Y},\mathcal{T})= 2 \partial^2_\mathcal{X}
%%\frac{\partial^2}{\partial \mathcal{X}^2}
%\ln F(\mathcal{X},\mathcal{Y},\mathcal{T}),
%\\
%F \equiv
%1+\exp(\mathcal{Z}_1)+\exp(\mathcal{Z}_2)
%+\exp(\mathcal{Z}_1+\mathcal{Z}_2+A_{12}),
%\\
%%\mathcal{Z}_i \equiv \kappa_i \left[ \mathcal{X}+ \lambda_i \mathcal{Y} -
%%\left(4\kappa_i^2+3\lambda_i^2 \right) \mathcal{T} +\delta_i \right],
%\exp(A_{12})=\frac{4(\kappa_1-\kappa_2)^2-(\lambda_1-\lambda_2)^2}
%{4(\kappa_1+\kappa_2)^2-(\lambda_1-\lambda_2)^2}
%\end{gather}
%\end{subequations}
%%
%
\begin{subequations}
\begin{gather}
U= 2 \partial^2_\mathcal{X}\ln\left( 1+e^{\mathcal{Z}_1}+e^{\mathcal{Z}_2}
+e^{\mathcal{Z}_1+\mathcal{Z}_2+A_{12}}\right),
%
%1+\exp(\mathcal{Z}_1)+\exp(\mathcal{Z}_2)
%+\exp(\mathcal{Z}_1+\mathcal{Z}_2+A_{12}),
\\
%%\mathcal{Z}_i \equiv \kappa_i \left[ \mathcal{X}+ \lambda_i \mathcal{Y} -
%%\left(4\kappa_i^2+3\lambda_i^2 \right) \mathcal{T} +\delta_i \right],
\exp(A_{12})=\frac{4(\kappa_1-\kappa_2)^2-(\lambda_1-\lambda_2)^2}
{4(\kappa_1+\kappa_2)^2-(\lambda_1-\lambda_2)^2},
\end{gather}
\end{subequations}
where $\mathcal{Z}_i \equiv \kappa_i \left[ \mathcal{X}+ \lambda_i \mathcal{Y} -
\left(4\kappa_i^2+3\lambda_i^2 \right) \mathcal{T} +\delta_i \right]$.

\begin{figure}[tbp]
\centering
\includegraphics[scale=0.37]{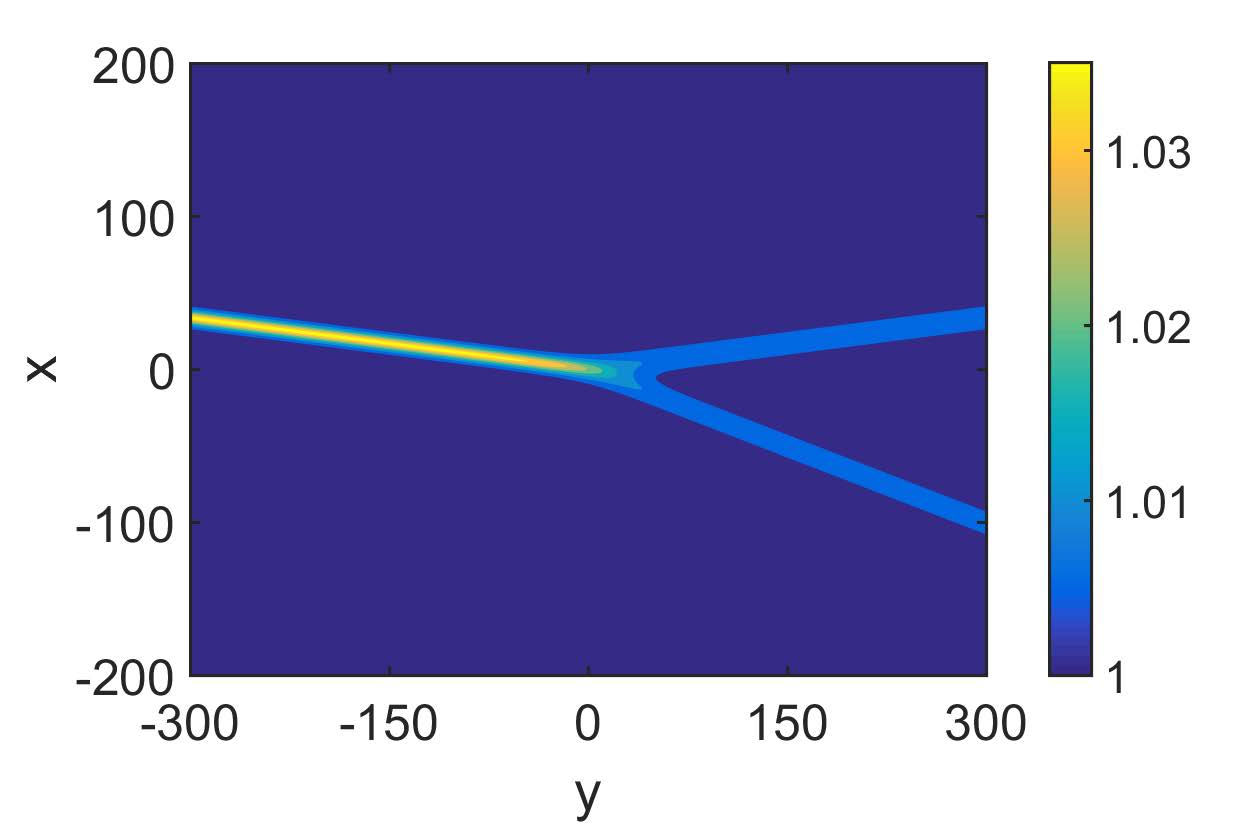}
\caption{A Y-type interaction for $2\kappa_1=\kappa_2=1$
and $\lambda_1=3\lambda_2=\frac{1}{4}$.}
\label{y}
\end{figure}

As was shown and observed in the context of shallow water waves \cite{PRE},
%
%While it is interesting on its own right that, depending on the sign of $\alpha$,
%two type of solitons may exist, i.e. dark (KPI) and anti-dark (KPII), their interactions
%may produce interesting patterns that also have been observed in water. In fact,
%it has been observed \cite{PRE} that
%
when two line solitons of the KPII intersect, a plethora of patterns can emerge.
We focus here on the ones most frequently observed in shallow waters.
To do this, fix $\epsilon=0.2$, $d^2=1/3$ and $u_0=1$ and choose two line solitons with
specific parameters, so that the angle of interaction will lead to different patterns. We evolve
these initial waves %for
up to $t=600$ so that they have enough time to interact. In Fig.~\ref{y} we show
the resonant interaction of two line solitons resulting into a Y-type wave. The parameters leading
to these interactions are summarized in the figure captions.

%Perhaps more interesting are the so-called X-type waves shown in Fig.~\ref{x}.

\begin{figure}[tbp]
\centering
\includegraphics[scale=0.37]{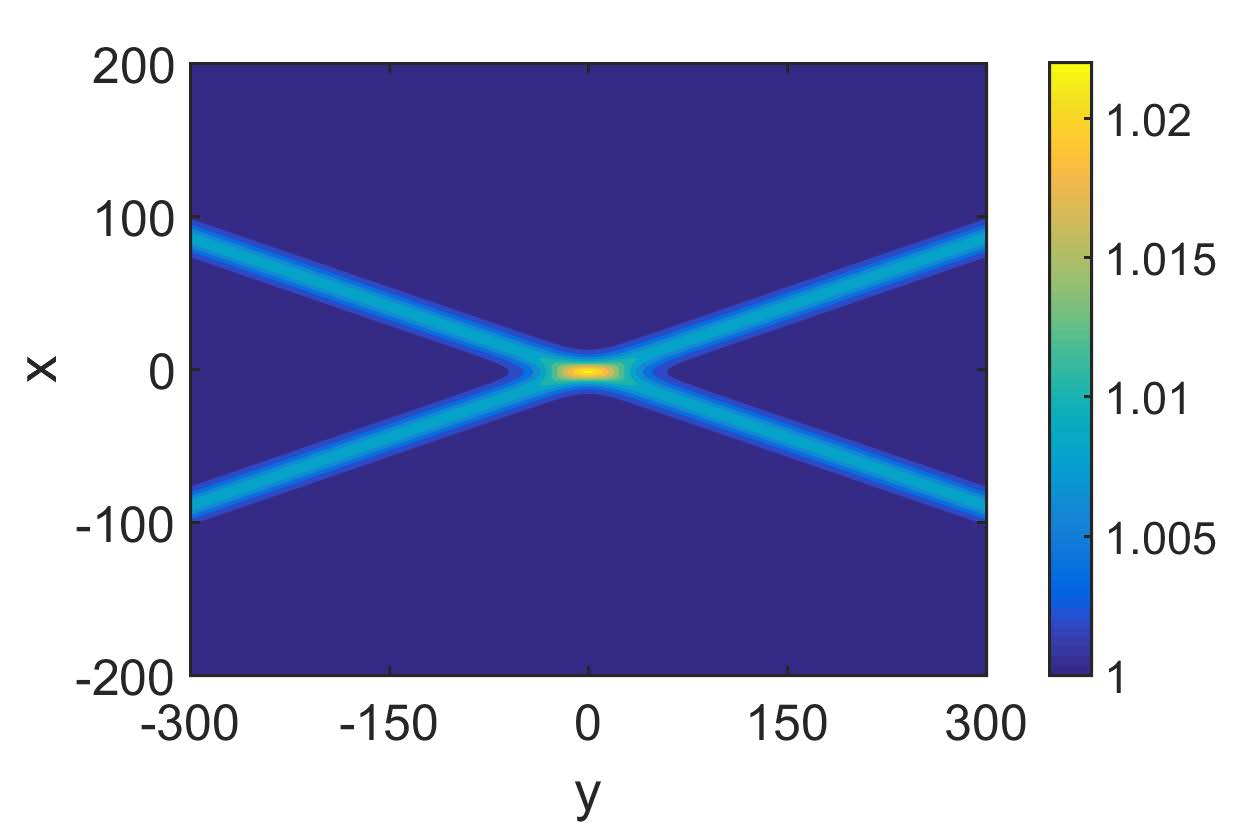}
\includegraphics[scale=0.37]{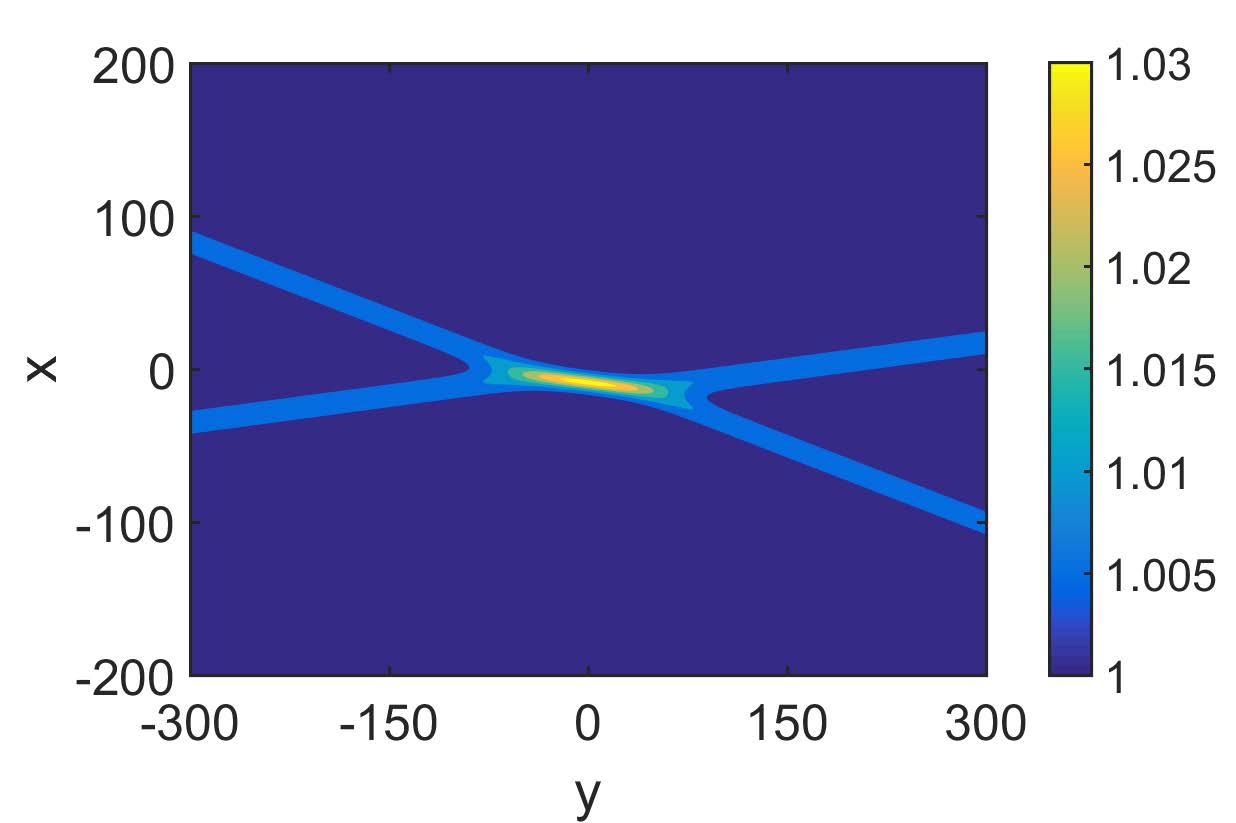}
\includegraphics[scale=0.37]{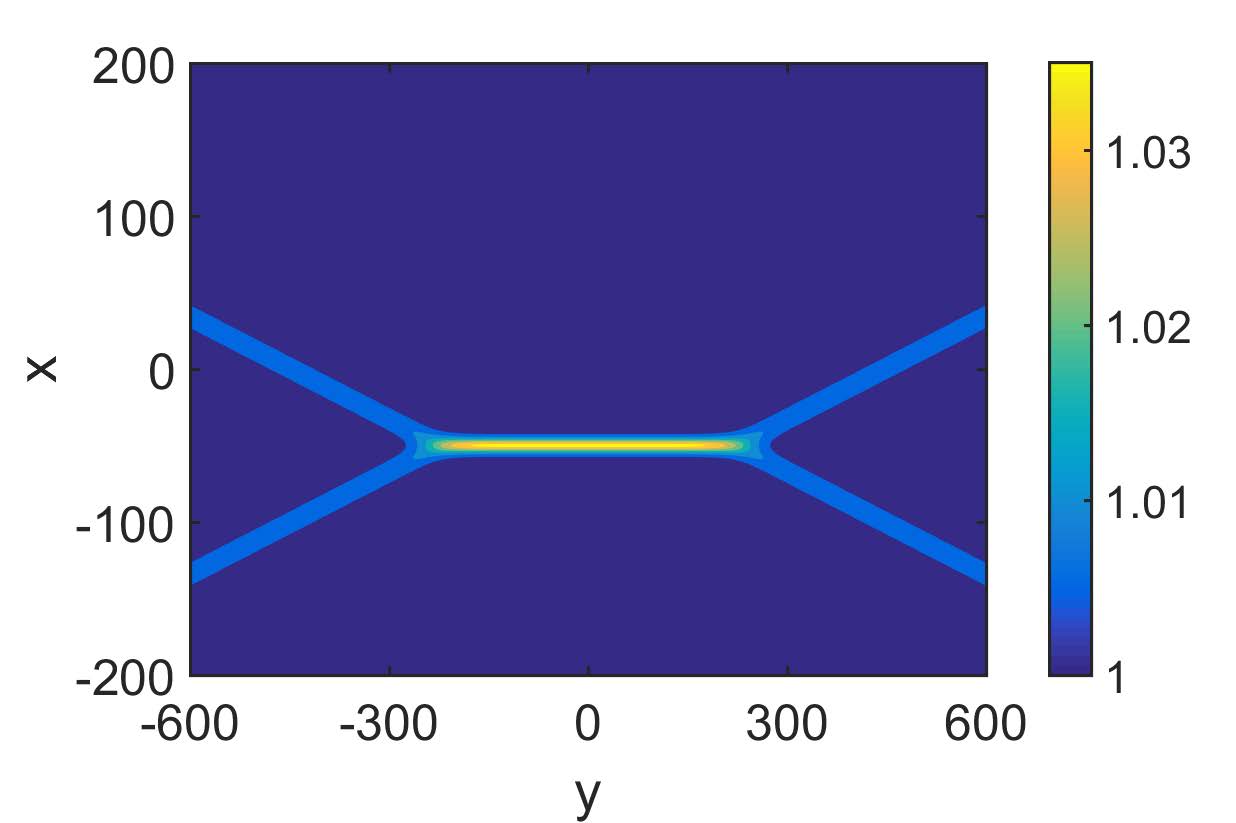}
\caption{(Color online) %Top: An
X-type interactions. Top: short stem,
%where
for $\kappa_1=\kappa_2=\frac{1}{2}$
and $\lambda_1=-\lambda_2=\frac{2}{3}$. Middle:
%An X-type interaction with longer
intermediate stem, for
%where
$\kappa_1=\kappa_2=\frac{1}{2}$, $\lambda_1=-\frac{1}{4}-10^{-2}$, and $\lambda_2=\frac{3}{4}$.
Bottom:
%An X-type interaction with a rather longer
long stem, for
%where
$\kappa_1=\kappa_2=\frac{1}{2}$ and $\lambda_1=-\lambda_1+10^{-10}=\frac{1}{2}$.}
\label{x}
\end{figure}

\begin{figure}[tbp]
\centering
\includegraphics[scale=0.37]{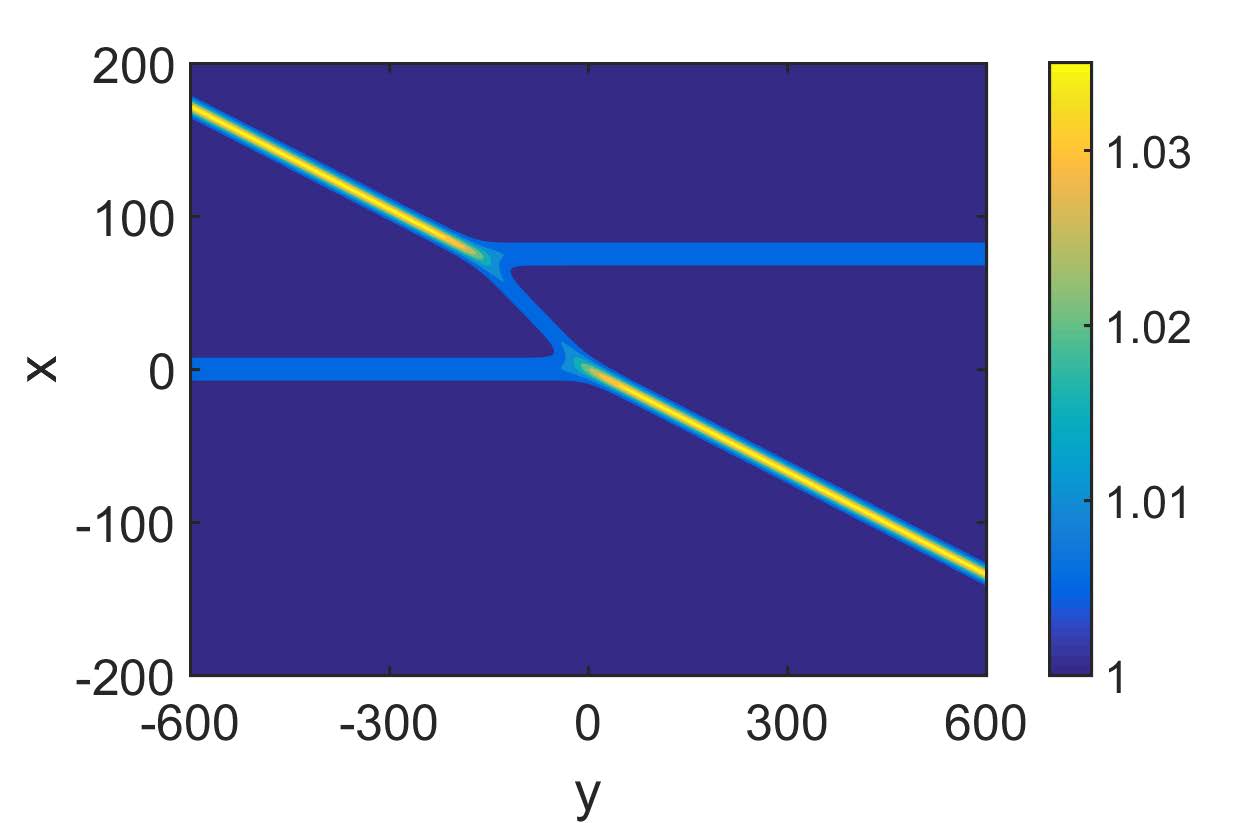}
\caption{An H-type interaction with $2\kappa_1=\kappa_2=1$,
%and
$\lambda_1=\frac{1}{2}-10^{-7}$, and $\lambda_2=0$.}
\label{h}
\end{figure}

X-type waves can also emerge, as shown in Fig.~\ref{x}.
These structures
%The X-type waves
are essentially discriminated by their ``stems'':
a short, intermediate and a long stem are respectively depicted in the top, middle
and bottom panels of Fig.~\ref{x}; notice that
%
%a short stem is
%depicted in Fig.~\ref{x}-top, an intermediate stem in Fig.~\ref{x}-middle, and a long stem in
%Fig.~\ref{x}-bottom, where
the long stem's height is higher than that of the incoming line solitons. In addition,
we can produce long stem interactions where the stem height is lower than the tallest incoming line
soliton, cf. Fig.~\ref{h}. We refer to these patterns as H-type.

%It should also be noted that
While we chose the above patterns as they
%seem to
appear more frequently in water,
%waves,
other
%novel and
more exotic, web-like structures
%have also been
are also supported by the KPII
%system
equation \cite{sarby3,sarby2,sarby1}, and
%can
may --in principle-- also be produced in optics.
%, as per our analysis.
%We will explore these patterns in a future communication.
Furthermore, these solutions, while approximate, also hold
well beyond the small-amplitude limit: similar results (not shown here) were obtained
even for $\epsilon =0.7$, with
the only additional effect being emission of noticeable radiation.
This is due to the robustness of
%because of the stability of the
KPII solitons, which is also verified by the observation
%emergence --and observation--
of these patterns
%results in these patterns to appear --and be observed--
in shallow water, even after the waves break \cite{PRE}.
%As such, we believe they can be also be
%experimentally realized in the same manner in optical media.

From the viewpoint of experiments, observations of antidark solitons
were reported in Refs.~\cite{segev2,Tang:16}. The Y-, X- and H-waves
may be observed experimentally using a setup similar to that of Ref.~\cite{segev2}.
In particular, one may employ at first a cw laser beam, which is split into two parts
via a beam-splitter. One branch goes through a cavity system
to form a pulse (as happens in typical pulsed lasers); this pulse branch
undergoes phase-engineering, i.e., passes through a phase mask so that
the characteristic phase jump of the antidark soliton is inscribed. Then,
the cw and the phase-engineered pulse are incoherently coupled inside the
nonlocal medium, e.g., a nematic liquid crystal, described by Eqs.~(\ref{NLS1})-(\ref{NLS2}).
This process forms one antidark soliton, as in Ref.~\cite{segev2}. To observe
Y-, X- or H-patterns predicted above, two such antidark solitons have to be
combined inside the crystal. The angle between the two incident beams, which
should be appropriately chosen so that a specific pattern be formed, can
be controlled by a rotating mirror in one of the branches.

\begin{figure}[ht]
\centering
\includegraphics[scale=0.37]{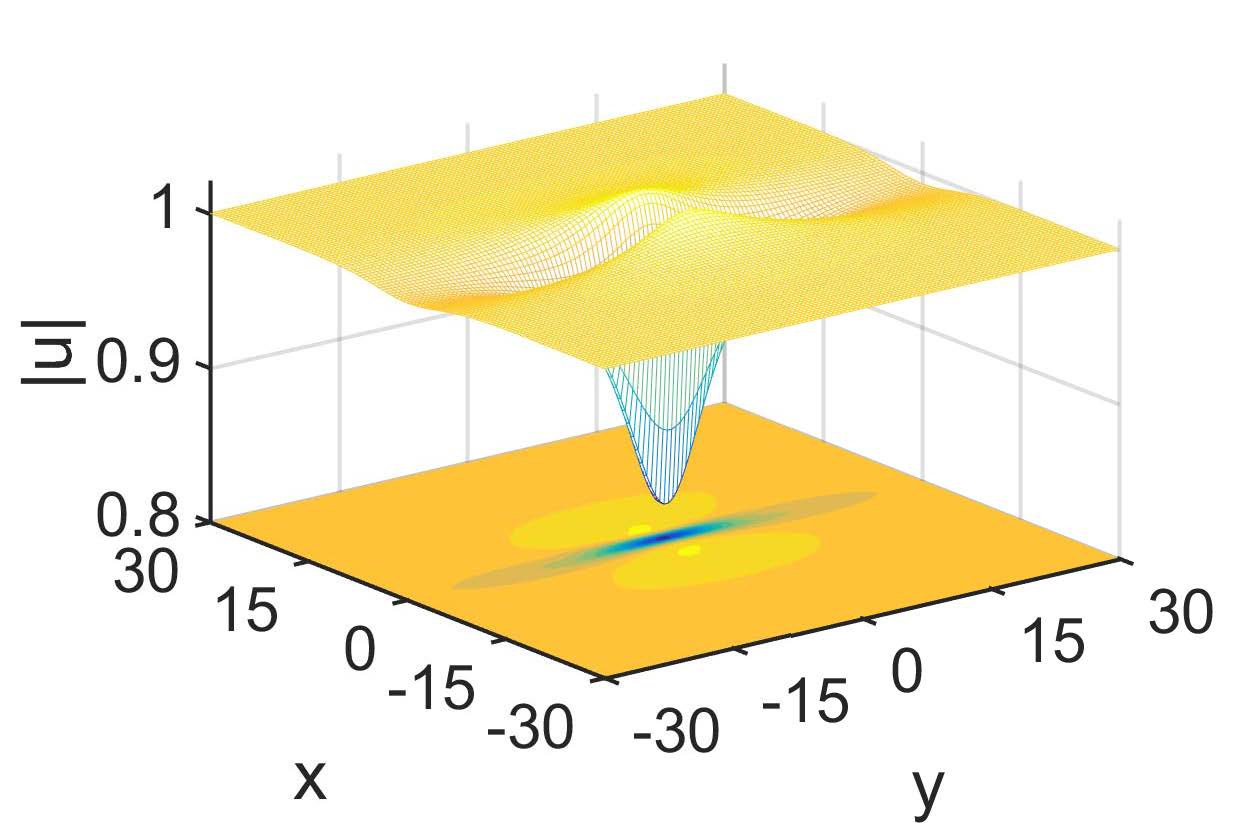}
\caption{A typical lump solution with $a=0$, $b=1$.}
\label{lump}
\end{figure}

Finally, let us consider the KPI case ($\alpha>0$). KPI also exhibits line soliton solutions, as
above, which are however unstable; it is thus most known for its
%lump
solution that decays algebraically in both spatial coordinates, i.e., the lump:
\begin{gather}
U(\mathcal{X},\mathcal{Y},\mathcal{T}) =
\nonumber \\
4 \frac{-(\mathcal{X} + a \mathcal{Y} + 3(a^2 - b^2)\mathcal{T})^2 + b^2 (\mathcal{Y}+6a \mathcal{T})^2
+ 1/b^2}
{[(\mathcal{X} + a \mathcal{Y} + 3(a^2 - b^2)\mathcal{T})^2 + b^2 (\mathcal{Y}+6a \mathcal{T})^2 + 1/b^2]^2}, \nonumber
\end{gather}
where $a,b$ are free real parameters.
%This solution has
Lumps have not yet been observed in water.
In Fig.~\ref{lump}, we show
%the result of
a direct simulation for the dark lump, and we
refer the reader to the recent work~\cite{baronio.kp} for
%further
details on multi-lump solutions and their interactions.

Concluding, we have established a homeomorphism between nonlocal nonlinear media
and shallow water waves. In particular, we have shown that there exists a surface
tension analogue in optics, depending on the nonlocality strength. This was identified
from the linear theory and was rigorously analyzed in the fully nonlinear regime,
by the asymptotic reduction of a nonlocal NLS system to the KP equations.
We demonstrated that, by depending on the effective surface tension, i.e., the
degree of nonlocality, novel structures can appear in optical nonlocal media. Thus,
fascinating phenomena that appear in water waves can also be observed in optics;
an example studied here is the emergence of X-, H-, or Y-shaped waves resulting
from the resonant interactions of stable line antidark solitons that were found to
exist in strongly nonlocal media. For weakly nonlocal ones, dark lumps were also
predicted to occur. The structures predicted in this work may in principle
be experimentally observed, either by choosing a material of specific nonlocality
or in a specific material altering the magnitude of the cw background, in a setup
similar to the one used for he observation of antidark solitons
\cite{segev2}.
%,Tang:16}.

\begin{acknowledgments}
We thank M. J. Ablowitz, P. J. Ioannou and E. P. Fitrakis for many useful discussions.
\end{acknowledgments}

%\bibliography{biblio_kp}

\begin{thebibliography}{29}
\expandafter\ifx\csname natexlab\endcsname\relax\def\natexlab#1{#1}\fi
\expandafter\ifx\csname bibnamefont\endcsname\relax
  \def\bibnamefont#1{#1}\fi
\expandafter\ifx\csname bibfnamefont\endcsname\relax
  \def\bibfnamefont#1{#1}\fi
\expandafter\ifx\csname citenamefont\endcsname\relax
  \def\citenamefont#1{#1}\fi
\expandafter\ifx\csname url\endcsname\relax
  \def\url#1{\texttt{#1}}\fi
\expandafter\ifx\csname urlprefix\endcsname\relax\def\urlprefix{URL }\fi
\providecommand{\bibinfo}[2]{#2}
\providecommand{\eprint}[2][]{\url{#2}}

\bibitem[{\citenamefont{Ablowitz}(Cambridge, 2011)}]{MJA1}
\bibinfo{author}{\bibfnamefont{M.~J.} \bibnamefont{Ablowitz}},
  \emph{\bibinfo{title}{Nonlinear dispersive waves: {A}symptotic analysis and
  solitons}} (\bibinfo{publisher}{Cambridge University Press},
  \bibinfo{year}{Cambridge, 2011}).

\bibitem[{\citenamefont{Zakharov and Kuznetsov}(1986)}]{zakharov}
\bibinfo{author}{\bibfnamefont{V.~E.} \bibnamefont{Zakharov}} \bibnamefont{and}
  \bibinfo{author}{\bibfnamefont{E.~A.} \bibnamefont{Kuznetsov}},
  \bibinfo{journal}{Physica D} \textbf{\bibinfo{volume}{18}},
  \bibinfo{pages}{455} (\bibinfo{year}{1986}).

\bibitem[{\citenamefont{Dauxois and Peyrard}(Cambridge, 2006)}]{dp}
\bibinfo{author}{\bibfnamefont{T.}~\bibnamefont{Dauxois}} \bibnamefont{and}
  \bibinfo{author}{\bibfnamefont{M.}~\bibnamefont{Peyrard}},
  \emph{\bibinfo{title}{Physics of Solitons}} (\bibinfo{publisher}{Cambridge
  University Press}, \bibinfo{year}{Cambridge, 2006}).

\bibitem[{\citenamefont{Ablowitz and Clarkson}(Cambridge, 1991)}]{BlackBook}
\bibinfo{author}{\bibfnamefont{M.~J.} \bibnamefont{Ablowitz}} \bibnamefont{and}
  \bibinfo{author}{\bibfnamefont{P.~A.} \bibnamefont{Clarkson}},
  \emph{\bibinfo{title}{Solitons, nonlinear evolution equations and inverse
  scattering}} (\bibinfo{publisher}{Cambridge University Press},
  \bibinfo{year}{Cambridge, 1991}).

\bibitem[{\citenamefont{Hasegawa and Kodama}(Oxford, 1995)}]{hasko}
\bibinfo{author}{\bibfnamefont{A.}~\bibnamefont{Hasegawa}} \bibnamefont{and}
  \bibinfo{author}{\bibfnamefont{Y.}~\bibnamefont{Kodama}},
  \emph{\bibinfo{title}{Solitons in optical communications}}
  (\bibinfo{publisher}{Claredon Press}, \bibinfo{year}{Oxford, 1995}).

\bibitem[{\citenamefont{Ablowitz and Baldwin}(2012)}]{PRE}
\bibinfo{author}{\bibfnamefont{M.~J.} \bibnamefont{Ablowitz}} \bibnamefont{and}
  \bibinfo{author}{\bibfnamefont{D.~E.} \bibnamefont{Baldwin}},
  \bibinfo{journal}{Phys. Rev. E} \textbf{\bibinfo{volume}{86}},
  \bibinfo{pages}{036305} (\bibinfo{year}{2012}).

\bibitem[{\citenamefont{Rotschild et~al.}(2005)\citenamefont{Rotschild, Cohen,
  Manela, Segev, and Carmon}}]{rot}
\bibinfo{author}{\bibfnamefont{C.}~\bibnamefont{Rotschild}},
  \bibinfo{author}{\bibfnamefont{O.}~\bibnamefont{Cohen}},
  \bibinfo{author}{\bibfnamefont{O.}~\bibnamefont{Manela}},
  \bibinfo{author}{\bibfnamefont{M.}~\bibnamefont{Segev}}, \bibnamefont{and}
  \bibinfo{author}{\bibfnamefont{T.}~\bibnamefont{Carmon}},
  \bibinfo{journal}{Phys. Rev. Lett.} \textbf{\bibinfo{volume}{95}},
  \bibinfo{pages}{213904} (\bibinfo{year}{2005}).

\bibitem[{\citenamefont{Krolikowski et~al.}(2004)\citenamefont{Krolikowski,
  Bang, Nikolov, Neshev, Wyller, Rasmussen, and Edmundson}}]{krol1}
\bibinfo{author}{\bibfnamefont{W.}~\bibnamefont{Krolikowski}},
  \bibinfo{author}{\bibfnamefont{O.}~\bibnamefont{Bang}},
  \bibinfo{author}{\bibfnamefont{N.~I.} \bibnamefont{Nikolov}},
  \bibinfo{author}{\bibfnamefont{D.}~\bibnamefont{Neshev}},
  \bibinfo{author}{\bibfnamefont{J.}~\bibnamefont{Wyller}},
  \bibinfo{author}{\bibfnamefont{J.~J.} \bibnamefont{Rasmussen}},
  \bibnamefont{and}
  \bibinfo{author}{\bibfnamefont{D.}~\bibnamefont{Edmundson}},
  \bibinfo{journal}{J. Opt. B: Quantum Semiclass. Opt.}
  \textbf{\bibinfo{volume}{6}}, \bibinfo{pages}{S288} (\bibinfo{year}{2004}).

\bibitem[{\citenamefont{Litvak et~al.}(1975)\citenamefont{Litvak, Mironov,
  Fraiman, and Yunakovskii}}]{litvak}
\bibinfo{author}{\bibfnamefont{A.~G.} \bibnamefont{Litvak}},
  \bibinfo{author}{\bibfnamefont{V.~A.} \bibnamefont{Mironov}},
  \bibinfo{author}{\bibfnamefont{G.~M.} \bibnamefont{Fraiman}},
  \bibnamefont{and} \bibinfo{author}{\bibfnamefont{A.~D.}
  \bibnamefont{Yunakovskii}}, \bibinfo{journal}{Sov. J. Plasma Phys.}
  \textbf{\bibinfo{volume}{1}}, \bibinfo{pages}{60} (\bibinfo{year}{1975}).

\bibitem[{\citenamefont{Yakimenko et~al.}(2005)\citenamefont{Yakimenko,
  Zaliznyak, and Kivshar}}]{plasma}
\bibinfo{author}{\bibfnamefont{A.~I.} \bibnamefont{Yakimenko}},
  \bibinfo{author}{\bibfnamefont{Y.~A.} \bibnamefont{Zaliznyak}},
  \bibnamefont{and} \bibinfo{author}{\bibfnamefont{Y.~S.}
  \bibnamefont{Kivshar}}, \bibinfo{journal}{Phys. Rev. E}
  \textbf{\bibinfo{volume}{71}}, \bibinfo{pages}{065603(R)}
  (\bibinfo{year}{2005}).

\bibitem[{\citenamefont{Conti et~al.}(2003)\citenamefont{Conti, Peccianti, and
  Assanto}}]{ass0}
\bibinfo{author}{\bibfnamefont{C.}~\bibnamefont{Conti}},
  \bibinfo{author}{\bibfnamefont{M.}~\bibnamefont{Peccianti}},
  \bibnamefont{and} \bibinfo{author}{\bibfnamefont{G.}~\bibnamefont{Assanto}},
  \bibinfo{journal}{Phys. Rev. Lett.} \textbf{\bibinfo{volume}{91}},
  \bibinfo{pages}{073901} (\bibinfo{year}{2003}).

\bibitem[{\citenamefont{Assanto}(2012)}]{ass1}
\bibinfo{author}{\bibfnamefont{G.}~\bibnamefont{Assanto}},
  \emph{\bibinfo{title}{Nematicons: {S}patial Optical Solitons in Nematic
  Liquid Crystals}} (\bibinfo{publisher}{New Jersey: Wiley-Blackwell},
  \bibinfo{year}{2012}).

\bibitem[{\citenamefont{Lahaye et~al.}(2009)\citenamefont{Lahaye, Menotti,
  Santos, Lewenstein, and Pfau}}]{dipolar}
\bibinfo{author}{\bibfnamefont{T.}~\bibnamefont{Lahaye}},
  \bibinfo{author}{\bibfnamefont{C.}~\bibnamefont{Menotti}},
  \bibinfo{author}{\bibfnamefont{L.}~\bibnamefont{Santos}},
  \bibinfo{author}{\bibfnamefont{M.}~\bibnamefont{Lewenstein}},
  \bibnamefont{and} \bibinfo{author}{\bibfnamefont{T.}~\bibnamefont{Pfau}},
  \bibinfo{journal}{Rep. Prog. Phys.} \textbf{\bibinfo{volume}{72}},
  \bibinfo{pages}{126401} (\bibinfo{year}{2009}).

\bibitem[{\citenamefont{Benney and Luke}(1964)}]{BL}
\bibinfo{author}{\bibfnamefont{D.~J.} \bibnamefont{Benney}} \bibnamefont{and}
  \bibinfo{author}{\bibfnamefont{J.~C.} \bibnamefont{Luke}},
  \bibinfo{journal}{J. Math. and Phys.} \textbf{\bibinfo{volume}{43}},
  \bibinfo{pages}{309} (\bibinfo{year}{1964}).

\bibitem[{\citenamefont{Coskun et~al.}(2000)\citenamefont{Coskun,
  Christodoulides, Kim, Chen, Soljacic, and Segev}}]{segev2}
\bibinfo{author}{\bibfnamefont{T.~H.} \bibnamefont{Coskun}},
  \bibinfo{author}{\bibfnamefont{D.~N.} \bibnamefont{Christodoulides}},
  \bibinfo{author}{\bibfnamefont{Y.-R.} \bibnamefont{Kim}},
  \bibinfo{author}{\bibfnamefont{Z.}~\bibnamefont{Chen}},
  \bibinfo{author}{\bibfnamefont{M.}~\bibnamefont{Soljacic}}, \bibnamefont{and}
  \bibinfo{author}{\bibfnamefont{M.}~\bibnamefont{Segev}},
  \bibinfo{journal}{Phys. Rev. Lett.} \textbf{\bibinfo{volume}{84}},
  \bibinfo{pages}{2374} (\bibinfo{year}{2000}).

\bibitem[{\citenamefont{Kivshar and Agrawal}(2003)}]{kivshar_book}
\bibinfo{author}{\bibfnamefont{Y.~S.} \bibnamefont{Kivshar}} \bibnamefont{and}
  \bibinfo{author}{\bibfnamefont{G.~P.} \bibnamefont{Agrawal}},
  \emph{\bibinfo{title}{Optical Solitons: From Fibers to Photonic Crystals}}
  (\bibinfo{publisher}{Academic Press}, \bibinfo{year}{2003}).

\bibitem[{\citenamefont{Ghofraniha et~al.}(2007)\citenamefont{Ghofraniha,
  Conti, Ruocco, and Trillo}}]{liq1}
\bibinfo{author}{\bibfnamefont{N.}~\bibnamefont{Ghofraniha}},
  \bibinfo{author}{\bibfnamefont{C.}~\bibnamefont{Conti}},
  \bibinfo{author}{\bibfnamefont{G.}~\bibnamefont{Ruocco}}, \bibnamefont{and}
  \bibinfo{author}{\bibfnamefont{S.}~\bibnamefont{Trillo}},
  \bibinfo{journal}{Phys. Rev. Lett.} \textbf{\bibinfo{volume}{99}},
  \bibinfo{pages}{043903} (\bibinfo{year}{2007}).

\bibitem[{\citenamefont{Conti et~al.}(2009)\citenamefont{Conti, Fratalocchi,
  Peccianti, Ruocco, and Trillo}}]{liq3}
\bibinfo{author}{\bibfnamefont{C.}~\bibnamefont{Conti}},
  \bibinfo{author}{\bibfnamefont{A.}~\bibnamefont{Fratalocchi}},
  \bibinfo{author}{\bibfnamefont{M.}~\bibnamefont{Peccianti}},
  \bibinfo{author}{\bibfnamefont{G.}~\bibnamefont{Ruocco}}, \bibnamefont{and}
  \bibinfo{author}{\bibfnamefont{S.}~\bibnamefont{Trillo}},
  \bibinfo{journal}{Phys. Rev. Lett.} \textbf{\bibinfo{volume}{102}},
  \bibinfo{pages}{083902} (\bibinfo{year}{2009}).

\bibitem[{\citenamefont{Zakharov and Kuznetsov}(1997)}]{zakh}
\bibinfo{author}{\bibfnamefont{V.~E.} \bibnamefont{Zakharov}} \bibnamefont{and}
  \bibinfo{author}{\bibfnamefont{E.~A.} \bibnamefont{Kuznetsov}},
  \bibinfo{journal}{Physics Uspekhi} \textbf{\bibinfo{volume}{40}},
  \bibinfo{pages}{1087} (\bibinfo{year}{1997}).

\bibitem[{\citenamefont{Kuznetsov et~al.}(1986)\citenamefont{Kuznetsov,
  Rubenchik, and Zakharov}}]{zakhpr}
\bibinfo{author}{\bibfnamefont{E.~A.} \bibnamefont{Kuznetsov}},
  \bibinfo{author}{\bibfnamefont{A.~M.} \bibnamefont{Rubenchik}},
  \bibnamefont{and} \bibinfo{author}{\bibfnamefont{V.~E.}
  \bibnamefont{Zakharov}}, \bibinfo{journal}{Phys. Rep.}
  \textbf{\bibinfo{volume}{142}}, \bibinfo{pages}{103} (\bibinfo{year}{1986}).

\bibitem[{\citenamefont{Infeld et~al.}(1994)\citenamefont{Infeld, Senatorski,
  and Skorupski}}]{infeld}
\bibinfo{author}{\bibfnamefont{E.}~\bibnamefont{Infeld}},
  \bibinfo{author}{\bibfnamefont{A.}~\bibnamefont{Senatorski}},
  \bibnamefont{and} \bibinfo{author}{\bibfnamefont{A.~A.}
  \bibnamefont{Skorupski}}, \bibinfo{journal}{Phys. Rev. Lett.}
  \textbf{\bibinfo{volume}{72}}, \bibinfo{pages}{1345} (\bibinfo{year}{1994}).

\bibitem[{\citenamefont{Kuznetsov and Turitsyn}(1982)}]{kuz1}
\bibinfo{author}{\bibfnamefont{E.~A.} \bibnamefont{Kuznetsov}}
  \bibnamefont{and} \bibinfo{author}{\bibfnamefont{S.~K.}
  \bibnamefont{Turitsyn}}, \bibinfo{journal}{JETP}
  \textbf{\bibinfo{volume}{55}}, \bibinfo{pages}{844} (\bibinfo{year}{1982}).

\bibitem[{\citenamefont{Pelinovsky et~al.}(1995)\citenamefont{Pelinovsky,
  Stepanyants, and Kivshar}}]{peli1}
\bibinfo{author}{\bibfnamefont{D.~E.} \bibnamefont{Pelinovsky}},
  \bibinfo{author}{\bibfnamefont{Y.~A.} \bibnamefont{Stepanyants}},
  \bibnamefont{and} \bibinfo{author}{\bibfnamefont{Y.~S.}
  \bibnamefont{Kivshar}}, \bibinfo{journal}{Phys. Rev. E}
  \textbf{\bibinfo{volume}{51}}, \bibinfo{pages}{5016} (\bibinfo{year}{1995}).

\bibitem[{\citenamefont{Kivshar and Pelinovsky}(2000)}]{pelirev}
\bibinfo{author}{\bibfnamefont{Y.~S.} \bibnamefont{Kivshar}} \bibnamefont{and}
  \bibinfo{author}{\bibfnamefont{D.~E.} \bibnamefont{Pelinovsky}},
  \bibinfo{journal}{Phys. Rep.} \textbf{\bibinfo{volume}{331}},
  \bibinfo{pages}{117} (\bibinfo{year}{2000}).

\bibitem[{\citenamefont{Chakravarty and Kodama}(2008)}]{sarby3}
\bibinfo{author}{\bibfnamefont{S.}~\bibnamefont{Chakravarty}} \bibnamefont{and}
  \bibinfo{author}{\bibfnamefont{Y.}~\bibnamefont{Kodama}},
  \bibinfo{journal}{J. Phys. A: Math. Theor.} \textbf{\bibinfo{volume}{41}},
  \bibinfo{pages}{275209} (\bibinfo{year}{2008}).

\bibitem[{\citenamefont{Chakravarty and Kodama}(2009)}]{sarby2}
\bibinfo{author}{\bibfnamefont{S.}~\bibnamefont{Chakravarty}} \bibnamefont{and}
  \bibinfo{author}{\bibfnamefont{Y.}~\bibnamefont{Kodama}},
  \bibinfo{journal}{Stud. Appl. Math.} \textbf{\bibinfo{volume}{123}},
  \bibinfo{pages}{83} (\bibinfo{year}{2009}).

\bibitem[{\citenamefont{Chakravarty et~al.}(2010)\citenamefont{Chakravarty,
  Lewkow, and Maruno}}]{sarby1}
\bibinfo{author}{\bibfnamefont{S.}~\bibnamefont{Chakravarty}},
  \bibinfo{author}{\bibfnamefont{T.}~\bibnamefont{Lewkow}}, \bibnamefont{and}
  \bibinfo{author}{\bibfnamefont{K.-I.} \bibnamefont{Maruno}},
  \bibinfo{journal}{Appl. Anal.} \textbf{\bibinfo{volume}{89}},
  \bibinfo{pages}{529} (\bibinfo{year}{2010}).

\bibitem[{\citenamefont{Tang et~al.}(2016)\citenamefont{Tang, Guo, Xiang, Shao,
  Song, Zhao, and Shen}}]{Tang:16}
\bibinfo{author}{\bibfnamefont{D.~Y.} \bibnamefont{Tang}},
  \bibinfo{author}{\bibfnamefont{J.}~\bibnamefont{Guo}},
  \bibinfo{author}{\bibfnamefont{Y.~J.} \bibnamefont{Xiang}},
  \bibinfo{author}{\bibfnamefont{G.~D.} \bibnamefont{Shao}},
  \bibinfo{author}{\bibfnamefont{Y.~F.} \bibnamefont{Song}},
  \bibinfo{author}{\bibfnamefont{L.~M.} \bibnamefont{Zhao}}, \bibnamefont{and}
  \bibinfo{author}{\bibfnamefont{D.~Y.} \bibnamefont{Shen}}, in
  \emph{\bibinfo{booktitle}{Photonics and Fiber Technology 2016 (ACOFT, BGPP,
  NP)}} (\bibinfo{publisher}{Optical Society of America},
  \bibinfo{year}{2016}), p. \bibinfo{pages}{JM6A.17}.

\bibitem[{\citenamefont{Baronio et~al.}(2016)\citenamefont{Baronio, Wabnitz,
  and Kodama}}]{baronio.kp}
\bibinfo{author}{\bibfnamefont{F.}~\bibnamefont{Baronio}},
  \bibinfo{author}{\bibfnamefont{S.}~\bibnamefont{Wabnitz}}, \bibnamefont{and}
  \bibinfo{author}{\bibfnamefont{Y.}~\bibnamefont{Kodama}},
  \bibinfo{journal}{Phys. Rev. Lett.} \textbf{\bibinfo{volume}{116}},
  \bibinfo{pages}{173901} (\bibinfo{year}{2016}).

\end{thebibliography}

\end{document}